# Effects of Compression on the Local Iodine Environment in Dipotassium Zinc Tetraiodate(V) Dihydrate K$_2$Zn(IO$_3$)$_4$·2H$_2$O

Daniel Errandonea,[1,*] Robin Turnbull,[1] Hussien H. H. Osman,[1,2] Zoulikha Hebboul[3], Pablo Botella[1], Neha Bura[1], Peijie Zhang[1], Jose Luis Rodrigo Ramon[1], Josu Sanchez-Martin[1], Catalin Popescu[4], Francisco J. Manjón[2]

[1]Departamento de Física Aplicada-ICMUV-MALTA Consolider Team, Universitat de Valencia, 46100 Valencia, Spain

[2]Instituto de Diseño para la Fabricación y Producción Automatizada, MALTA Consolider Team, Universitat Politècnica de València, 46022 València, Spain

[3]Laboratoire Physico-Chimie des Matériaux, Université Amar Telidji de Laghouat, BP 37G, Route de Ghardaia, Laghouat 03000, Algeria

[4]CELLS-ALBA Synchrotron Light Facility, Cerdanyola 08290, Barcelona, Spain; https://orcid.org/0000-0001-6613-4739; Email: cpopescu@cells.es

*Corresponding authors, Email: daniel.errandonea@uv.es

**Abstract:** Combining X-ray diffraction with density-functional theory and electron topology calculations we found that pressure substantially modifies the bonding in $K_2Zn(IO_3)_4 \cdot 2H_2O$. We discovered that under compression there is a progressive change from primary covalent I–O bonds and secondary halogen I···O interactions towards O–I–O electron-deficient multicenter bonds. Because of this, iodine hypercoordination converts $IO_3$ trigonal pyramids towards $IO_6$ units. The formation of these $IO_6$ units breaks the typical isolation of iodate molecules forming an infinite two-dimensional iodate network. Hypercoordination influences the hydrogen atoms too, such that multicenter O–H–O bonds are also promoted with increasing pressure. We have determined that $K_2Zn(IO_3)_4 \cdot 2H_2O$ is one of the most compressible iodates studied to date, with a bulk modulus of 22(3) GPa. The pressure-induced structural changes strongly modify the electronic structure as shown by optical-absorption measurements and band-structure calculations. The band-gap energy closes from 4.2(1) eV at ambient pressure to 3.4(1) eV at 20 GPa.

KEYWORDS: iodate, high-pressure, XRD, crystal structure, DFT, bonding, band gap

## I. Introduction

Secondary non-covalent interactions have garnered significant attention across multiple domains of chemistry and physics. In particular, halogen bonding, which occurs as a secondary interaction involving halogen elements, e.g. between iodine and second-neighbor O atoms, are a prevalent and crucial form of non-covalent interactions in many compounds and have been the subject of extensive research for numerous years.[1] Of particular interest are iodate materials, which contain iodine with oxidation state 5+.[2] In multiple compounds, the $I^{5+}$ cation is found in a tri-coordinated arrangement, giving rise to iodate ions, $IO_3^-$, forming trigonal pyramids that exhibit a tetrahedral configuration typical of $sp^3$ hybridization when the fourth vertex, occupied by a lone electron pair (LEP), is considered. This configuration gives the $IO_3^-$ ion a permanent electric dipole moment, which results in notable and significant macroscopic anisotropic effects in iodate materials[2] Within the $IO_3^-$ pyramids, there are three primary short I–O bonds which are delocalized resonant (covalent-like) bonds. These I–O bonds are a mixture of single and double covalent bonds where there is a delocalization of electrons corresponding to the π bonds. In addition, there are also longer secondary non-covalent I⋯O interactions between neighboring iodate ions that have been traditionally considered as halogen bonds.

The investigation of the compressibility of the non-covalent interactions in response to external factors, such as elevated pressure, can deepen the comprehension of these interactions and aid in the design of compounds like metal iodates.[2] The utilization of high pressure to modify the atomic spacing within solids is a well-established technique and can lead to considerable modifications in the chemical and physical characteristics of materials. Given that iodates contain robust covalent I–O bonds alongside weaker halogen I⋯O interactions, investigations under high pressure have been especially beneficial in examining the properties of these compounds, such as the metal iodates.[2] One of the significant discoveries from the high-pressure investigations in iodates is that the application of external pressure results in an increase in the coordination number (hypercoordination) of the iodine atom. This is not surprising since one well-known rule of high-pressure science, known as the pressure-coordination rule, stablishes that in general, there is always an increase in atomic coordination with increasing pressure.[3] What is more surprising is that the combination of X-ray diffraction (XRD) measurements and computational simulations (also supported by Raman scattering measurements) in recent articles of iodates at high pressure has led to the conclusion that in many iodates the hypercoordination leads to the formation of electron-deficient multicenter bonds (EDMBs).[4] Such bonds are characterized by a combination of localized electrons, like those found in covalent materials, and delocalized electrons, akin to those in metals, and have recently been used to reinterpret the geometries of polyiodides[5] and to explain the bonding in phase-change materials.[6] The pressure-induced formation of three-center O–I–O EDMBs from short primary iono-covalent bonds, such as the I–O bonds in $IO_3^-$ units, and long secondary non-covalent interactions, such as the I⋯O halogen bonds between $IO_3^-$ units, is consistent with previous works proposing an unified theory of multicenter bonding.[7,8]

We have previously used high-pressure to study different iodates, like $Mg(IO_3)_2$,[9] and hydrated iodates, like $Ca(IO_3)_2·H_2O$ and $Ba(IO_3)_2·H_2O$.[10,11] Research has also been recently conducted on complex iodates, such as $Na_3Bi(IO_3)_6$,[12] on iodic acid $HIO_3$,[13] and more intricate systems like $Sr(IO_3)_2·HIO_3.3]$.[4] In all these systems a bonding transformation has been observed at pressures below 20 GPa. Dipotassium zinc tetraiodate(V) dihydrate, $K_2Zn(IO_3)_4·2H_2O$, and related compounds like $K_2Mn(IO_3)_4·2H_2O$, $K_2Co(IO_3)_4·2H_2O$, and $K_2Mg(IO_3)_4·2H_2O$ form another group of interesting compounds,[14] which contains I⋯O halogen bonds and O–H⋯O hydrogen bonds. However, they have never been studied under high-pressure conditions.

In this work, we report the study of $K_2Zn(IO_3)_4·2H_2O$ under high-pressure conditions. The characterization was performed using synchrotron XRD, optical absorption, and density-functional theory calculations. In addition, we have performed a theoretical analysis of the topology of the electron density to understand the change in chemical bonding as pressure increases. The present study brings to light the influence of pressure on the crystal structure, bonding, and electronic properties of $K_2Zn(IO_3)_4·2H_2O$ contributing to the understanding of the behavior of complex iodates under compression.

## II. Methods

### II.I. Experiments

Crystals of $K_2Zn(IO_3)_4 \cdot 2H_2O$ were prepared from zinc chloride ($ZnCl_2$, purity ≥98% from Aldrich) and potassium iodate ($KIO_3$, purity ≥99.5% from Aldrich) with a 1:4 molar ratio, which were mixed in a distilled water solution at room temperature. This procedure was followed by very slow evaporation. The obtained crystals were put a second time for 1 month into a solution of the same type as the original $ZnCl_2$ + 4 $KIO_3$ solution to promote the growth of the size of the crystals. The crystal structure was confirmed to be the monoclinic structure reported in the literature[14,15] The unit-cell parameters are $a$ = 13.804(9) Å, $b$ = 7.728(5) Å, $c$ = 8.286(6) Å, and $\beta$ = 126.6(1)° if the structure is described by space group $C2$[14] or alternatively by $a$ = 8.286(6) Å, $b$ = 7.728(5) Å, $c$ = 11.086 (8) Å, and $\beta$ = 90.3(1)° if the structure is described by space group $I2$.[15]

We performed two powder XRD experiments under HP, one up to 10.60(5) GPa (run 1) and another up to 20.10(5) GPa (run 2). We also carried out one optical absorption experiment up to 20.10(5) GPa. For all HP studies, we employed a membrane-driven diamond-anvil cell with anvils with a culet measuring 500 μm in diameter. Stainless-steel gaskets, with a thickness of 200 μm, pre-indented to a thickness of 55 μm, and with a centered hole of 170 μm, were employed. The pressure-transmitting medium consisted of a 4:1 mixture of methanol and ethanol, which facilitates quasi-hydrostatic conditions up to 10 GPa,[16] but has been used successfully to study iodates up to 20 GPa.[4,6] In the XRD experiments pressure was determined from the XRD pattern of copper grains loaded next to the sample using the equation of state (EoS) of copper reported by Dewaele *et al*. from XRD experiments.[17] In optical experiments pressure measurements were obtained through ruby fluorescence.[18] In both cases the accuracy was better than 0.05 GPa.

Synchrotron powder X-ray diffraction (XRD) experiments were conducted at the BL04-MSPD beamline of the ALBA synchrotron,[19] utilizing a monochromatic X-ray beam with a wavelength of 0.4246 Å. The X-ray beam was focused on a spot size of 20 μm × 20 μm. Data collection for XRD was performed using a Rayonix SX165 charge-coupled device. The resulting two-dimensional patterns were processed with Dioptas,[20] while FullProf software[21] was employed for the analysis, specifically for Rietveld refinement of the integrated one-dimensional XRD patterns. Optical absorption experiments were performed in the ultraviolet-visible-near-infrared range, utilizing an optical setup that included a deuterium lamp, reflecting optical objectives, and an Ocean Optics spectrometer.[22] The optical absorption was determined by dividing the transmittance spectrum of the sample at normal incidence by that of the reference source.

### II.II. Calculations

The electronic structure calculations for the $K_2Zn(IO_3)_4 \cdot 2H_2O$ system were performed within the framework of density functional theory (DFT), utilizing the Vienna *Ab initio* Simulation Package (VASP).[23-25] The exchange-correlation effects were treated using the generalized gradient approximation (GGA) as formulated by Perdew, Burke, and Ernzerhof (PBE),[26] in conjunction with the projector-augmented wave (PAW) method.[27] The revised PBEsol functional[28] along with the dispersion correction DFT-D3 method of Grimme with zero-damping function[29] were employed to describe the exchange and correlation energy. The Brillouin zone was sampled with Γ-centered Monkhorst–Pack meshes of 5 × 5 × 5,[30] and the plane-wave basis set was truncated at an energy cutoff ($E_{cutoff}$) of 850 eV to ensure total energy convergence within $10^{-5}$ eV per atom. The pseudopotentials include the ($3p^6$, $4s^1$), ($5s^2$, $5p^5$), ($3d^{10}$, $4s^2$), ($2s^2$, $2p^4$), and ($1s^1$) electrons treated as valence electrons for K, I Zn, O, and H atoms, respectively.

To examine the electronic density topology of the investigated solids, a density-based approach was employed within the framework of the Quantum Theory of Atoms in Molecules (QTAIM).[31] We have used this approach because it is rather easy to perform with the CRITIC2 program[32] of Otero de la Roza *et al.* which has shown similar results to DGrid software[33] as recently discussed.[34] For this purpose, single-point calculations were carried out using Quantum ESPRESSO (version 6.5),[35] in combination with Wannier90[36] and the CRITIC2 software.[32] These calculations were performed at the optimized geometries obtained from VASP, employing the same *k*-point grids. A plane-wave energy cutoff of 100 Ry and a charge density cutoff of 400 Ry were used consistently across all simulations. The pseudopotentials used to describe Kohn-Sham states, as well as PAW datasets for obtaining the all-electron density, were sourced from the pslibrary.[37] The delocalization index (DI), which provides a measure of the number of electrons shared (ES) via the relation ES = 2 × DI, was determined using a Wannier transformation, as described previously in the literature.[38] Crystal structures were visualized and analyzed using VESTA program.[39] The VASPKIT program[40] was utilized for various purposes dealing with the density of states (DOS) and band structures data obtained from the DFT calculations.

### III. Results and discussion

#### III.I. Influence of pressure in the crystal structure

$K_2Zn(IO_3)_4 \cdot 2H_2O$ has a monoclinic crystal structure described by space group *C*2 [13] or alternatively by space group *I*2 [14]. In this work, we will use the second option which has a monoclinic β angle close to 90º and is more convenient to describe the changes induced by pressure. The $K_2Zn(IO_3)_4 \cdot 2H_2O$ crystal structure is represented in Figure 1.

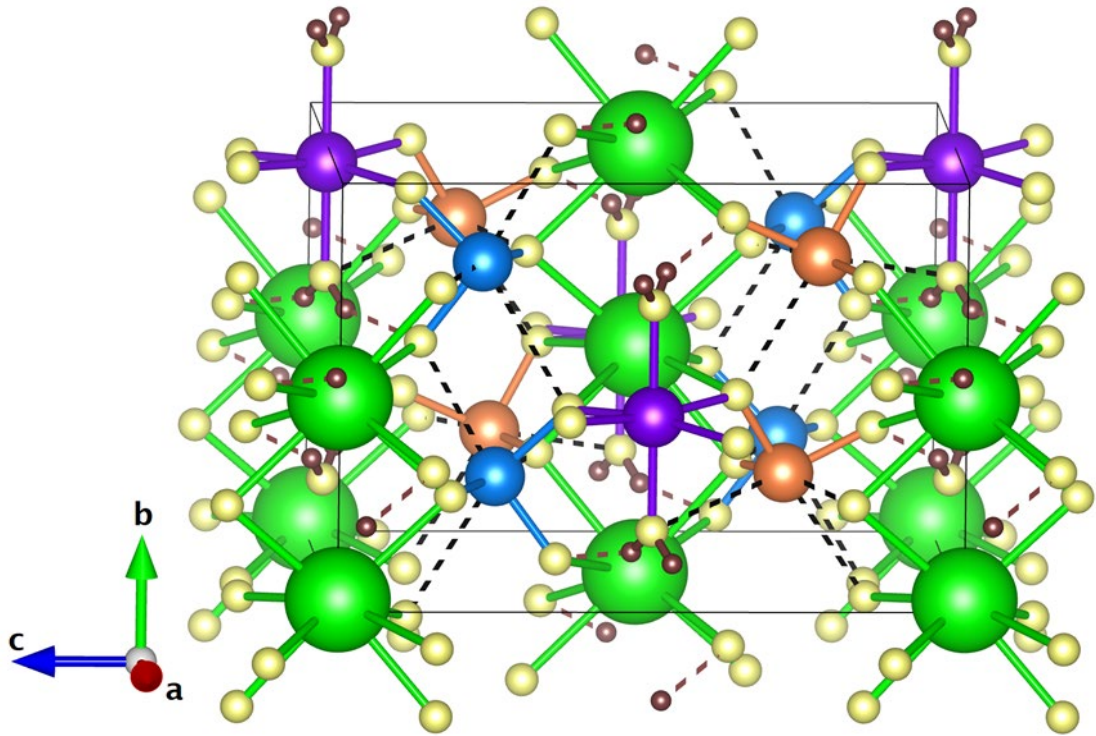

**Figure 1:** Crystal structure of $K_2Zn(IO_3)_4 \cdot 2H_2O$ represented using space group *I*2. K atoms are shown in green, Zn atoms in purple, O atoms in yellow, and H atoms in brown. Iodine atoms are located at two different Wyckoff positions in the crystal structure, and they are shown in blue (I1) and orange (I2). The coordination polyhedra are shown in the figure.

Halogen I···O bonds are shown with black dashed lines and O–H···O hydrogen bonds are shown with brown dashed lines.

The crystal structure has two symmetrically independent potassium (K) and iodine (I) atoms (which we shall name I1 and I2). At ambient pressure, these atoms, along with the zinc (Zn) atoms, are coordinated through shared oxygen (O) atoms. Additionally, the Zn atoms are coordinated by two water molecules positioned in a trans configuration. The K, Zn, and O atoms from the water molecule occupy special positions along two-fold symmetry axes. In the structure, there are also O–H···O hydrogen bonds and I···O halogen bonds. Both iodate atoms (I1 and I2) are arranged in a trans configuration around the Zn atoms. The iodine atoms are coordinated by three oxygen atoms forming trigonal $[IO_3]^-$ pyramids. They also form three I···O halogen bonds. In Figure 2(a) we show the local environment of $[IO_3]^-$ units. Assuming that K atoms act as cations and give their charge to other entities and considering the Zn coordination octahedron (linked to two $H_2O$ molecules and four $IO_3$ molecules), the formula unit can be interpreted as consisting of two $K^+$ cations and one $[Zn(IO_3)_4(H_2O)_2]^{2-}$ polyanion. In this light, an interesting characteristic of the crystal structure is that the packing of the $[Zn(IO_3)_4(H_2O)_2]^{2-}$ anions leads to the creation of one-dimensional channels running along the *b*-axis, which are occupied by $K^+$ ions (see Figure 2). Another curiosity is that the iodate molecules, due to the asymmetry in the electron distribution around the iodine atom caused by the LEP, are polarized. The polarizations of the $I2O_3^-$ groups nearly offset one another, while the polarizations of the $I1O_3^-$ groups are oriented in a parallel fashion. This alignment results in a net dipole moment directed along the *b*-axis. Readers should keep in mind that the description we provided above is for the crystal structure at ambient pressure, since the actual electron distribution in the solid might be pressure dependent as we will show in this work.

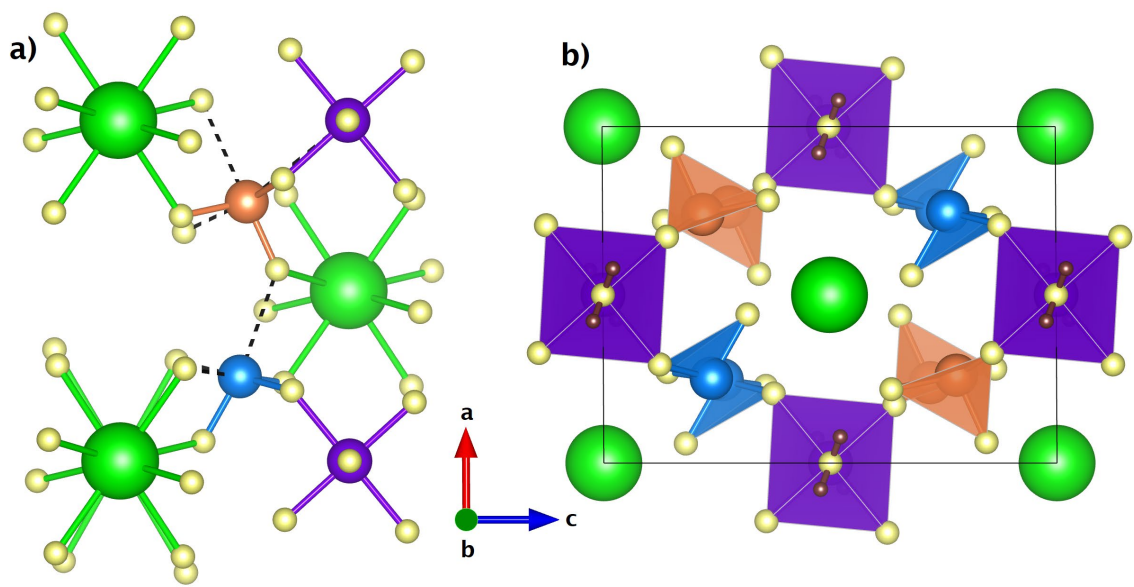

**Figure 2: (a)** Local environment of [IO$_3$]$^-$ units showing coordination polyhedra. **(b)** Projection of the crystal structure of K$_2$Zn(IO$_3$)$_4$·2H$_2$O along the *b* axis to emphasize the one-dimensional channels (corners and center) occupied by K atoms. Only the Zn and I coordination polyhedra are shown. I1 (I2) atoms are shown in blue (orange), K atoms a in green, Zn atoms in purple, O atoms in yellow, and H atoms in brown. The same projection is used in both figures.

To investigate the structural behavior of K$_2$Zn(IO$_3$)$_4$·2H$_2$O under compression, we conducted high-pressure X-ray diffraction experiments. The XRD patterns presented at selected pressures in Figure 3 demonstrate a continuous shift of the peaks towards higher angles, which is a direct result of lattice compression. There are two significant changes in XRD patterns. One is the splitting of peaks $\bar{2}02$ and 202, and peaks $\bar{2}24$ and 224 above 5 GPa. This fact demonstrates the increase of the monoclinic angle, $\beta$, under compression. Another noticeable fact is the increase of the separation between peaks 200 and 020 which indicates an anisotropic compression of the crystal structure. Notably, it shows that the *b*-axis is more compressible than the *a*-axis.

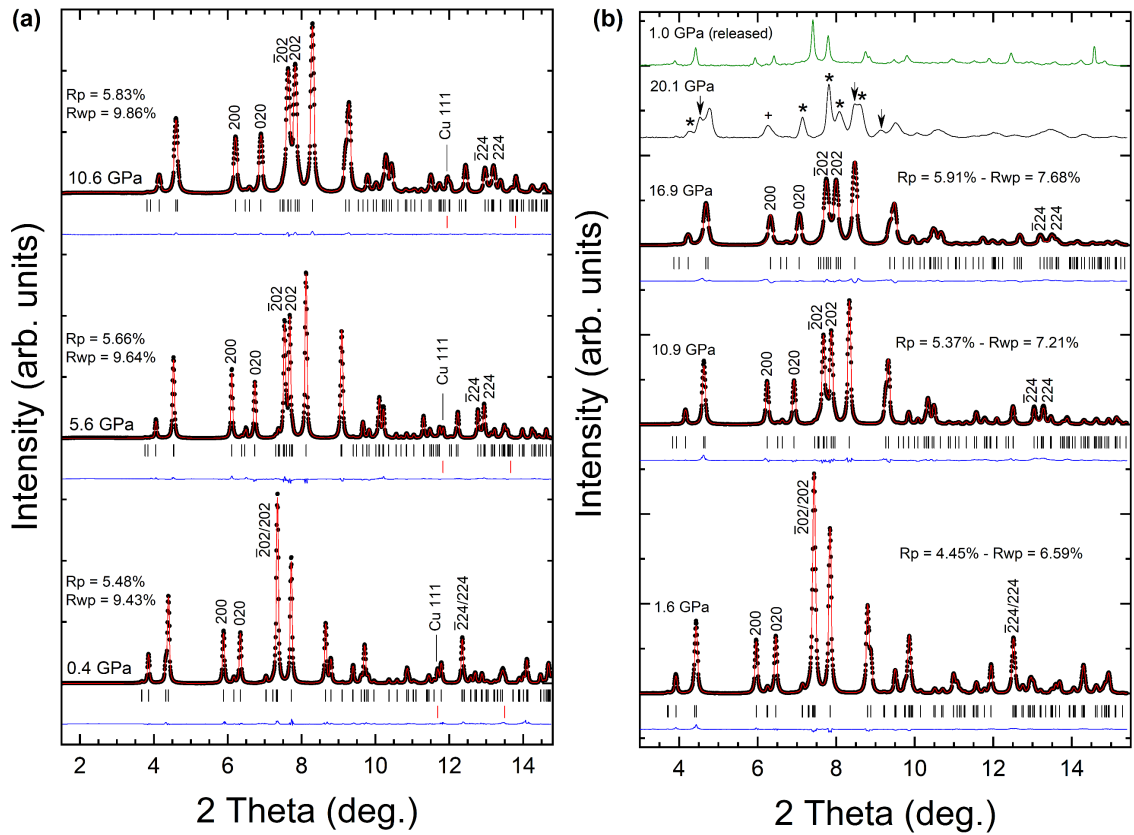

**Figure 3:** Selection of $K_2Zn(IO_3)_4 \cdot 2H_2O$ powder XRD patterns acquired at different pressures in the two experiments performed. Experimental data are shown with black circles and refinements with red lines. The residuals are shown with blue lines. Black ticks show the position of Bragg peaks. Reflections used in the discussion are identified with Miller indices. R-values of the refinements are given in the figure. In (a) we identify the Cu peak used to determine pressure and show the position of Cu peaks using red ticks. At the top of (b) we show an XRD pattern measured at 20.1 GPa with a black line, which shows evidence of the onset of a phase transition. Asterisks, arrows, and plus symbols identify the peaks discussed in the text. An XRD pattern measured after decompression is shown in (b) with a green line. The (r) next to pressure is used to distinguish pressures measured under pressure release.

We found that up to 19.55(5) GPa all peaks can be indexed according to the same structural model as the ambient pressure monoclinic structure (see Fig. 3). At 21.10(5) GPa, additional diffraction peaks emerge in the XRD patterns at 2θ = 4.5º, 8.5º, and 9.2º (they are indicated by arrows in the figure). We also found that the peak identified as 200 moves to lower angles at the same pressure (see the peak identified by the '+' symbol in the figure). The peak is at 6.33º at 16.90(5) GPa and at 6.25º at 21.10(5) GPa. Both facts indicate the onset of a phase transition. The remaining peaks in the pattern measured at 21.10(5) GPa correspond to the low-pressure phase (e.g. see those identified by asterisks in Fig. 3b).

The coexistence of both phases above the transition pressure implies that the phase transition is of first order in nature. The phase coexistence persists up to 25.05(5) GPa, the highest pressure reached in this study, so it has precluded the identification of the crystal structure of the HP phase, which remains an open issue for future studies. Furthermore, the changes observed in the XRD patterns are entirely reversible, as shown by the XRD pattern obtained during pressure release at 1.00(5) GPa and shown in green in Fig. 3b. This pattern resembles the pattern measured at 1.60(5) GPa before compression. All peaks in this pattern can be assigned to the low-pressure phase of $K_2Zn(IO_3)_4 \cdot 2H_2O$.

From the XRD data, we extracted the pressure dependence of the unit-cell parameters. The results are represented in Figure 4. The $\beta$ angle increases non-linearly from 90.3(1)º at ambient pressure to 91.6(1)º at 19.60(5) GPa, confirming the enhancement of the monoclinic distortion of the structure under compression. Fig. 4 shows that DFT calculations agree with experiments regarding the pressure dependence of the unit-cell parameters. The calculations present a slight underestimation of the lattice parameters *a* and *b*, as illustrated in Figure 4a. This discrepancy is typical of DFT calculations and results in a 7% underestimation in the estimated unit-cell volume, as shown in Figure 4b. Nevertheless, the calculations exhibit a comparable pressure dependence than that observed in experiments for all four lattice parameters.

The crystal structure of $K_2Zn(IO_3)_4 \cdot 2H_2O$ is monoclinic, therefore, the analysis of its compressibility requires the use of the monoclinic compressibility tensor which has four independent components.[41] The eigenvectors of this tensor give the main axes of compressibility of the crystal and the respective eigenvalues give their compressibility. We established them with the PASCAL tool[42] by using only the experimental data obtained from both runs 1 and 2 measured below 10 GPa, to ensure that they are not influenced by non-hydrostatic effects. The main compressibility axes are (010), (904), and ($\bar{3}$04). The corresponding compressibility values are 8.0(1) $10^{-3}$ GPa, 5.8(1) $10^{-3}$ GPa, and 4.0(1) $10^{-3}$ GPa. Consequently, the *b*- axis is the most compressible axis of K$_2$Zn(IO$_3$)$_4$·2H$_2$O. Its compressibility is 38% and 100% larger than the compressibility of the other two main axes of compressibility.

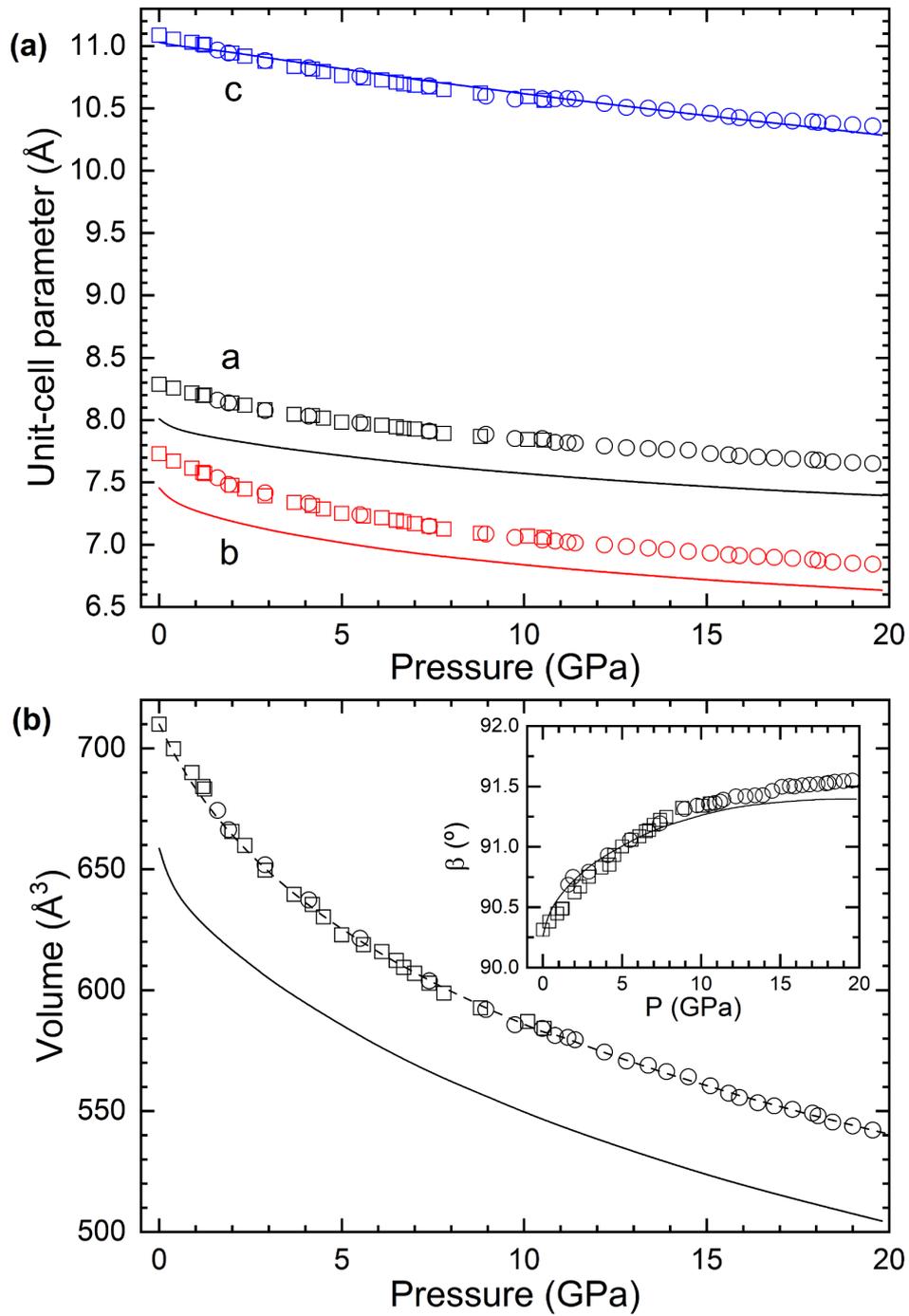

**Figure 4:** (a) Pressure dependence of the K$_2$Zn(IO$_3$)$_4$·2H$_2$O unit-cell parameters. (b) Pressure dependence of the K$_2$Zn(IO$_3$)$_4$·2H$_2$O unit-cell volume, with the inset showing the pressure dependence of the β angle. Results from run 1 (run 2) are shown as squares (circles). Solid lines represent the results of DFT calculations, while in (b) the dashed line represents the Birch-Murnaghan EoS fitted from experiments.

The alignment of the channels occupied by K$^+$ ions along the *b*-axis corresponds with the direction of maximum compressibility, explaining the observed anisotropic compressibility. The network of [Zn(IO$_3$)$_4$(H$_2$O)$_2$]$^{2-}$ anions is expected to be less compressible than KO$_8$ dodecahedra because K–O bonds can be assumed to be more compressible than Zn–O bonds. The reason for such an assumption is the inverse correlation between bond length and bond strength; shorter bonds are generally stronger than longer ones.[43] In this context, the average K–O bond distance at ambient pressure (2.81(7) Å) is longer than the average Zn–O bond distance at ambient pressure (2.12(3) Å). Therefore, it is reasonable to hypothesize that the KO$_8$ chains aligned along the *b*-axis make it the most compressible one.

The relationship between volume and pressure obtained from experiments was modeled using a third-order Birch-Murnaghan EoS.[44] This analysis yielded the unit-cell volume at zero pressure, $V_0$ = 710 (1) Å$^3$, the bulk modulus at zero pressure, $K_0$ = 22(3) GPa, and the pressure derivative of the bulk modulus, $K_0'$ = 10.2(9). The fitted EoS is shown as a dashed line in Fig. 4. From DFT calculations, we obtained $V_0$ = 653(3) Å$^3$, $K_0$ = 27(3) GPa, and $K_0'$ = 9(2). The bulk modulus and its pressure derivative agree within one standard deviation. These results indicate that, along with Zn(IO$_3$)$_2$, Mg(IO$_3$)$_2$, and Sr(IO$_3$)$_2$HIO$_3$,[4,6,45] K$_2$Zn(IO$_3$)$_4$·2H$_2$O is among the most compressible iodates examined to date (see Table 1).

**Table 1** - Compressibility data for highly compressible metal iodates.

| Sample | $V_0$ (Å$^3$) | $K_0$ (GPa) | $K_0'$ | Reference |
|---|---|---|---|---|
| K$_2$Zn(IO$_3$)$_4$·2H$_2$O | 710(3) | 22(3) | 10.2(9) | This work |
| Zn(IO$_3$)$_2$ | 265(1) | 21.6(7) | 7.0(3) | 45 |
| Mg(IO$_3$)$_2$ | 553(2) | 22.2(8) | 4.2(4) | 6 |
| Sr(IO$_3$)$_2$HIO$_3$ | 839(3) | 23(2) | 6.7(6) | 4 |

Typically, oxides exhibit a $K_0'$ value of around 4. The $K_0'$ value obtained in this work, 9(2) - 10.2(9), indicates that as pressure rises, the bulk modulus increases much faster than in most oxides. As observed in other iodates,[3,4] the observed phenomenon might be associated with the function of the iodine LEP, which facilitates the reduction of the long secondary I⋯O bond length under compression. The pressure-induced approach of the second neighboring oxygen atoms to the iodine atoms might cause the modification of the bonding of iodine, thereby triggering a rapid increase of the bulk modulus, which is reflected in the large value of its pressure derivative.

To better understand the effect of pressure on iodine bonding we have studied the pressure dependence of the I–O and I⋯O bonds. We determined the distances between iodine and oxygen atoms from Rietveld refinements and from DFT calculations. The results are represented in Figure 5. Calculations slightly overestimate the bond distances of the first neighbor covalent I–O bonds, which correspond to intramolecular bonds in the $IO_3$ unit and have a length of approximately 1.8 Å at ambient pressure. Additionally, calculations underestimate the length of secondary I⋯O bonds, which are intermolecular bonds between $IO_3$ units and are longer than 2.7 Å at ambient pressure. Despite these small differences, experiments and calculations show the same behavior under compression. The three short covalent bonds of the $IO_3$ pyramids become slightly enlarged under compression, while the long secondary I⋯O bonds suffer a notable decrease (up to 18 %) from 0 to 20 GPa.

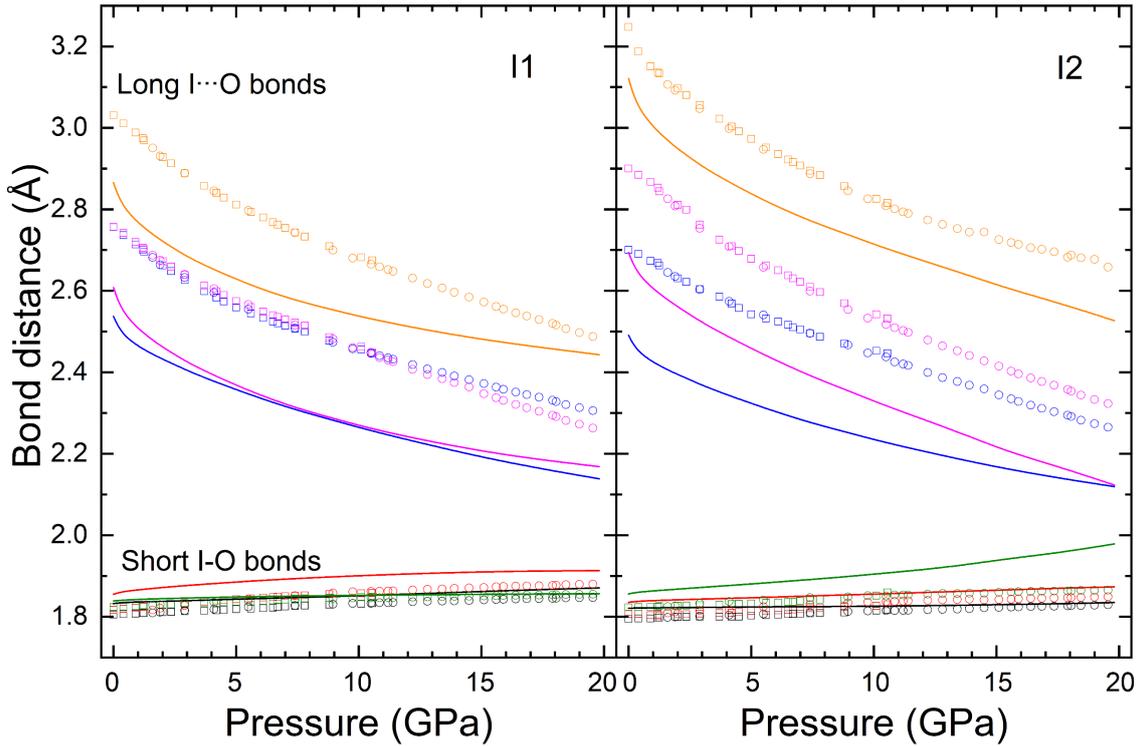

**Figure 5:** Pressure dependence of the experimental (symbols) and calculated (lines) bond distance between iodine and oxygen atoms for both I1 (left) and I2 (right) atoms in $K_2Zn(IO_3)_4 \cdot 2H_2O$.

At 20 GPa, the two closest O atoms to both I1 and I2 atoms from neighbor $IO_3$ molecules are within 2.25 – 2.35 Å, i.e. these long secondary bonds are only *ca*. 20% larger than the longest covalent bonds of the $IO_3$ molecule, which have value of 1.9 Å at 20 GPa. In addition, there is a sixth oxygen atom at 2.5 – 2.7 Å at 20 GPa. This implies a notable change in the coordination of iodine atoms. The close approach of two neighboring oxygen atoms to each iodine atom leads to a six-fold hypercoordination of I atoms and the formation of $IO_6$ trigonal bipyramids. In other words, the change in coordination of iodine leads to the formation of layers made of corner-sharing distorted $IO_6$ polyhedra, such as the one represented in Figure 6a. These layers run along the (001) plane and are separated by layers formed by $KO_8$ and $ZnO_6$ polyhedra. In addition to the layers one can observe that the I and O atoms along the *a*-axis form zigzag O–I–O chains (they are infinite –O1–I1–O2–I2–O1– chains) in which the O–I–O angle is close to 180° and the I–O–I angle is around 135° at 20 GPa (see Figure 6b).

The increase in bond length of the short I–O bonds observed in $K_2Zn(IO_3)_4 \cdot 2H_2O$ and other iodates under compression is related to the well-known paradox of high-pressure science, known as the pressure-distance paradox, which stablishes that despite the increase in atomic coordination with increasing pressure the bond distance could increase instead of decrease.[3] We could distinguish two types within this paradox: i) the increase in bond length on increasing pressure, as it happens in elemental Se and Te when the stable trigonal phase at room pressure is compressed, and ii) the increase in bond length on increasing pressure after a phase transition; e.g., the larger Si–O bond length in stishovite (with sixfold coordination for Si) than in coesite (with fourfold coordination for Si). Some of us have shown that the first type of this paradox is due to the formation of multicenter bonds and corresponds to the stage 2 of the process of multicenter bond formation. This is what happens in trigonal Se and Te as discussed in our previous work.[4] On the other hand, the second type of the pressure-distance paradox has been traditionally explained in the following way: the bond distance in the phase of higher atomic coordination must be higher than in the phase of lower atomic coordination to leave enough space for the new bonds to be formed. For example, in coesite, with four Si–O bonds, O atoms are 10.2% closer to the Si atom than in stishovite, with six Si–O bonds, because some free space must be left for the two new O atoms that come close to Si atom in the stishovite phase.

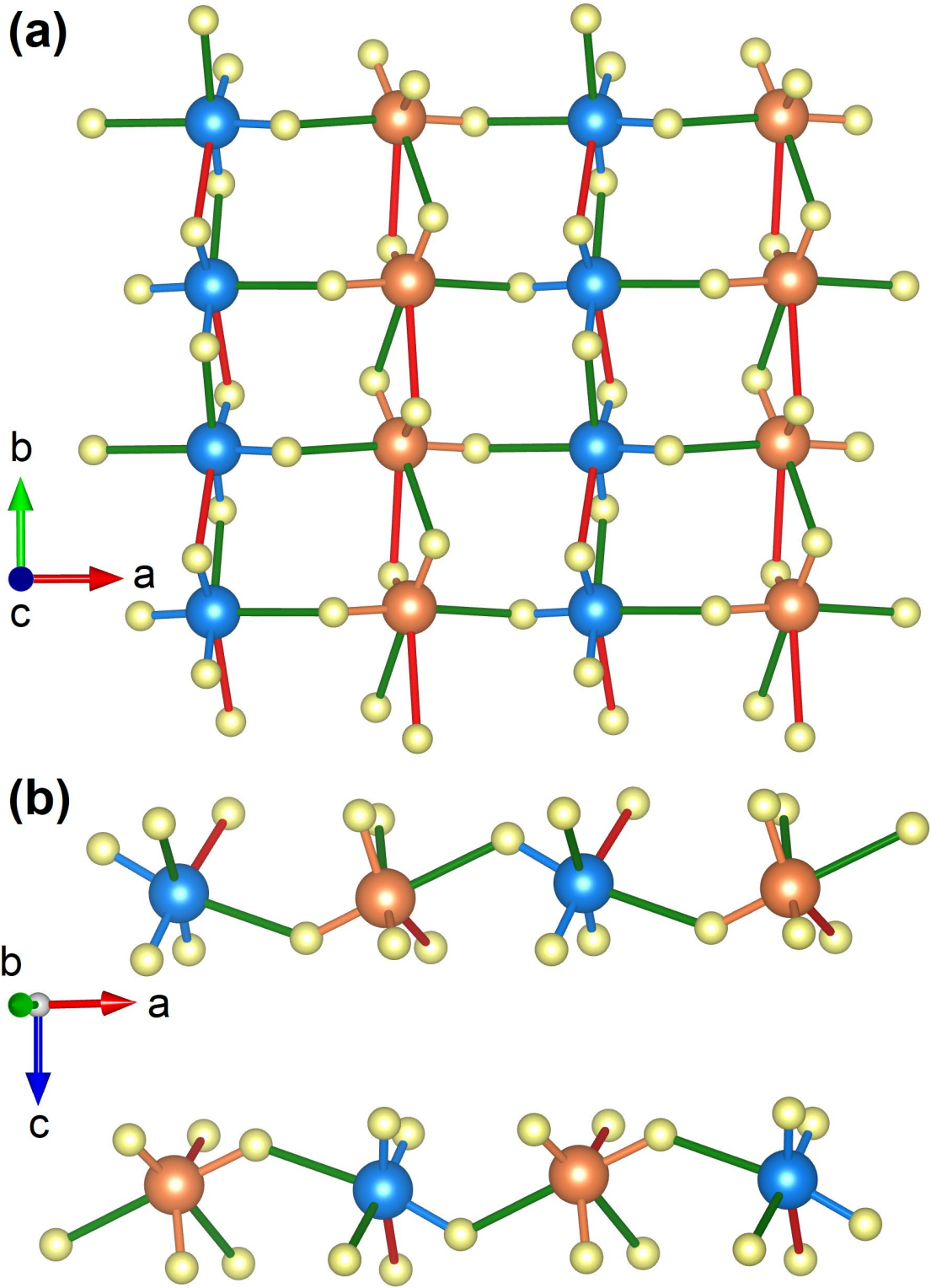

**Figure 6: (a)** Layers formed by hypercoordinated IO$_6$ molecules in K$_2$Zn(IO$_3$)$_4$·2H$_2$O at 20 GPa. I1 (I2) atoms are shown in blue (orange) and oxygen atoms in yellow. The bonds formed under compression are shown in green (2.25 – 2.35 Å) and red (2.5 – 2.7 Å). **(b)** The zigzag chain formed by the I-O bonds along the *a*-axis. Cations are located between layers.

In order to understand the change in chemical bonding in the I–O bonds with increasing pressure as the I atom becomes hypercoordinated, we have calculated the number of electrons shared (ES) between two atoms (I and O) for the short I–O and long I⋯O bonds (see details in section 2.b) and how they evolve with the change in the I–O and I⋯O bond lengths (see Figure 7). It can be observed that the ES value shows values between 2 and 2.25 for all short I1–O and I2–O bonds at room pressure. These high values confirm the covalent character of the I1–O and I2–O bonds within the $IO_3$ units. As pressure increases, there is a progressive decrease in ES value of the short I–O bonds and a concomitant increase of the ES value of the long I⋯O bonds. This behavior correlates with the increase of the short I–O bond distance and the decrease of the long I⋯O bond distances. Moreover, all short I–O and long I⋯O bonds show a tendency to reach the same bond distances and the same ES values at high pressures. A tendency that it is also reproduced by the short O-H covalent bonds and the long O⋯H (hydrogen) bonds (see bottom of Fig. 7).

The pressure dependence of the bond lengths and ES values in I–O bonds, as well as the gradual formation of linear or quasi-linear O–I–O bonds (as those shown in Figure 6b), is consistent with the pressure-induced formation of multicenter bonds, and in particular with the tendency to formation of electron-deficient multicenter bonds (EDMBs) since the charge gained by the long I⋯O bonds is at expense of the charge lost by the short I–O bonds, as explained by the proposed unified theory of multicenter bonding.[7,8] A simple way to illustrate the pressure-induced formation of EDMBs is to localize the different I–O bonds in the ES vs ET map on the basis of their ES and ET values at different pressures, where ET indicates the normalized number of electrons transferred between the two atoms as obtained from the Bader charges (see Figure 8). It can be observed how the average values of the I1–O and I2–O bonds decrease on going from 0 to 25 GPa, thus moving in the direction of formation of EDMBs (green region of the map). We must note that the formation of EDMBs is not completed even up to 25 GPa. In summary, our calculations confirm that there is a progressive change from covalent I–O bonds towards O–I–O EDMBs as pressure increases and iodine hypercoordination evolves from $IO_3$ units towards the regular $IO_6$ units. It must be stressed that the same tendency to hypercoordination and low ES values also affects the H atoms, so multicenter O–H–O bonds are also promoted with increasing pressure, in line with what is expected from the proposed unified theory of multicenter bonding.[7,8]

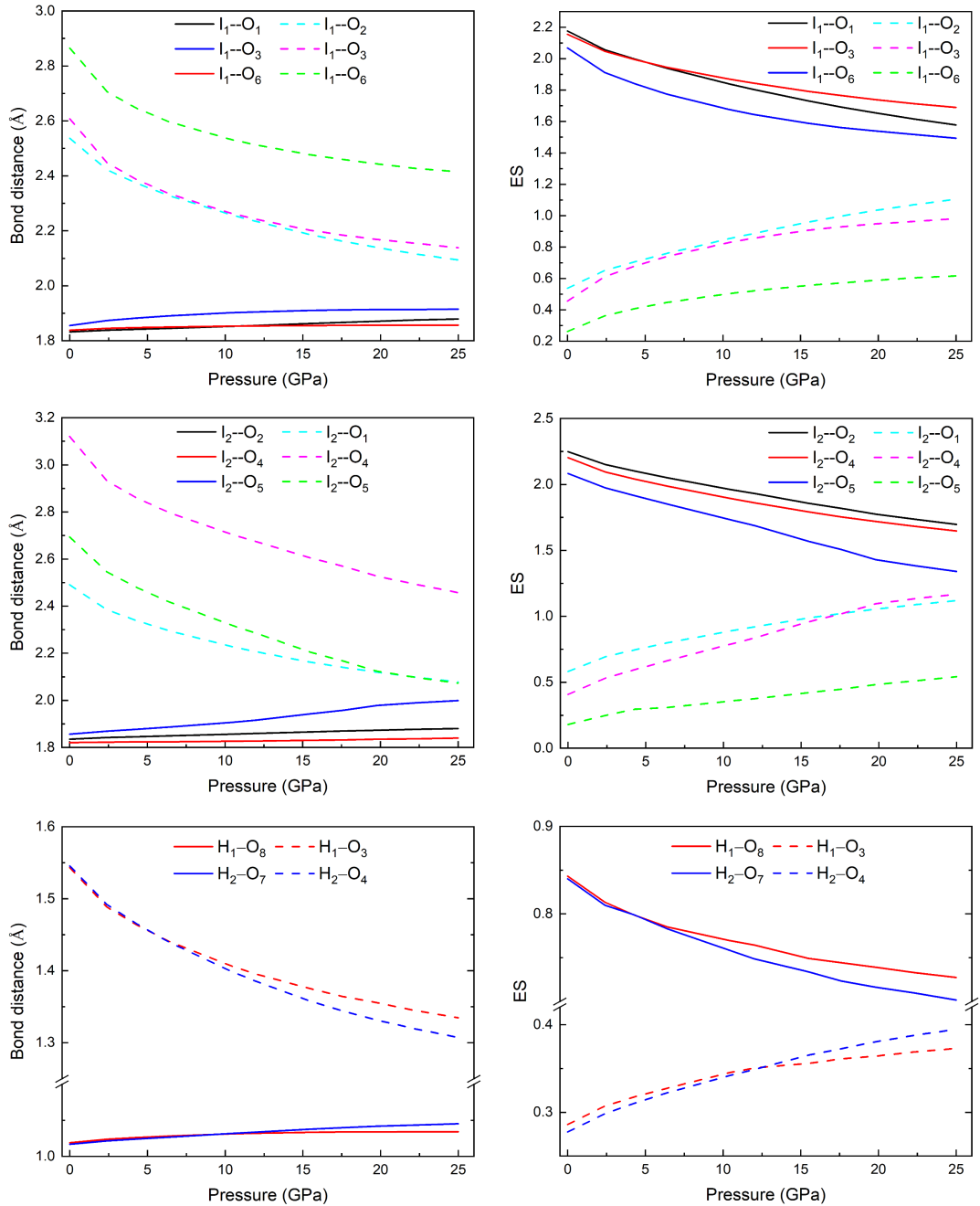

**Figure 7:** Theoretical pressure dependence of the I–O and H–O bond distances (left) and their corresponding ES values (right) in K$_2$Zn(IO$_3$)$_4$·2H$_2$O.

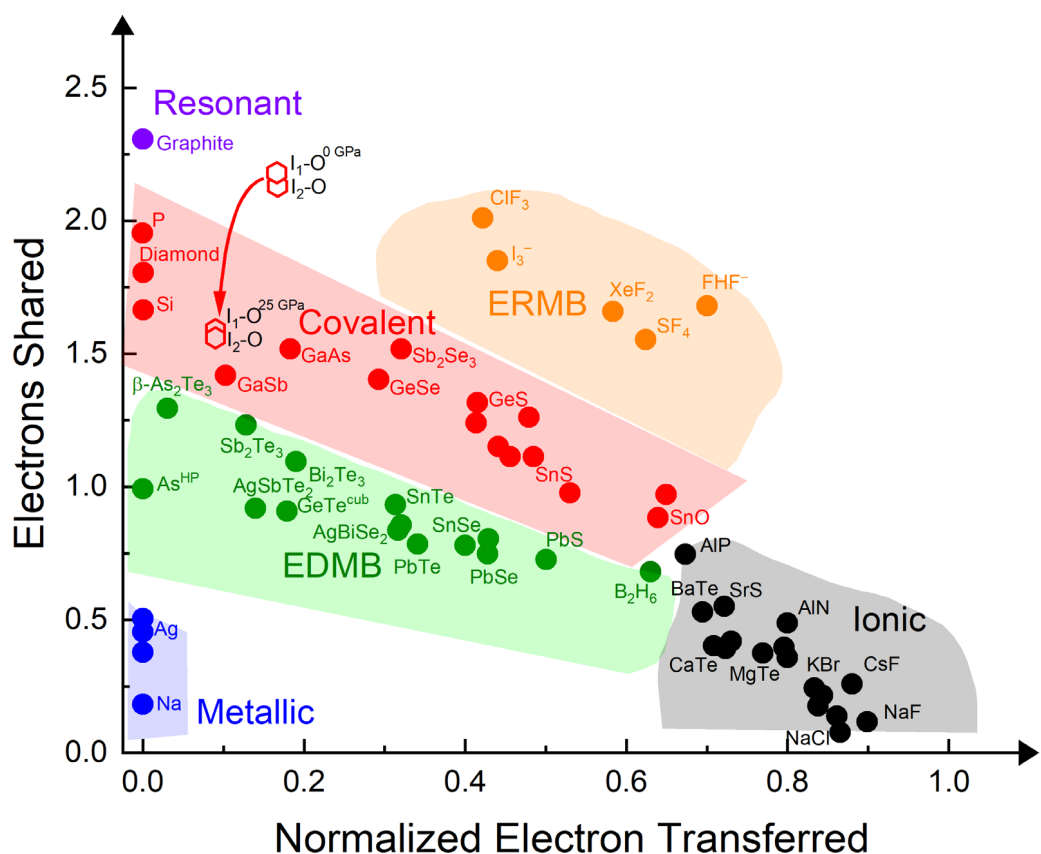

**Figure 8:** ES vs ET map showing the evolution of the I–O bonds in $K_2Zn(IO_3)_4 \cdot 2H_2O$. At 0 GPa all I–O bonds are in the covalent region.

It must be also mentioned that the pressure dependence of the bond lengths and ES values in H–O bonds also tend to the formation of linear or quasi-linear O–H–O bonds from the original covalent H–O bond and the O⋯H hydrogen bonds. It has been already suggested[46] that the strengthening of the hydrogen bond, to form a linear or quasi-linear symmetric O–H–O bond occurs when the O⋯H bond length decreases to a limit value of 1.2 Å, as obtained from an extensive series of O–H⋯O bonds.[47] A recent example seems to confirm this limit since the formation of the symmetric O–H–O bond in hydrogen pyrocarbonate shows a value of ca. 1.20 Å for the two symmetric O–H bond lengths of the O–H–O bond.[48] Note that our long O⋯H values in Figure 7 also show a tendency towards this limit value which is not attained at the pressure of 25 GPa, in which the long O⋯H bond length is still slightly above 1.3 Å. Therefore, our DFT values suggest the formation of linear or quasi-linear O–H–O multicenter bonds in this iodate under compression, whose proper study will require further work.

### III.II. Influence of pressure on the electronic structure

To determine the band-gap energy ($E_g$) and study the influence on pressure on the electronic structure of $K_2Zn(IO_3)_4 \cdot 2H_2O$, we carried out optical-absorption measurements and band-structure calculations. In Figure 9 we show the absorbance of $K_2Zn(IO_3)_4 \cdot 2H_2O$ at 0.30(5) GPa. The absorbance has three distinctive features: 1) A low-energy exponential absorption tail, which is typical of iodates,[3] and is related to defects and surface effects, and is commonly known as Urbach tail.[49] 2) A parabolic weak absorption of energies higher than 4.15 eV, which resembles an indirect band gap absorption.[49] 3) A sharp and strong absorption, which resembles a direct band gap absorption, at 4.5 eV. Considering the three contributions,[50] the absorbance of $K_2Zn(IO_3)_4 \cdot 2H_2O$ at 0.30(5) GPa can be well described, as shown in Figure 9. In the figure the Urbach tail is shown in orange, the indirect absorption in blue, and the direct absorption in green. The red line shows the overall fit of the absorbance spectrum considering the three aforementioned contributions. The conclusion that the absorption spectrum contains contributions from both direct and indirect absorptions is supported by the present band-structure calculations shown in Figure 10(a). According to calculations at 0 GPa, there is an indirect band gap between the $\Gamma$ point of the valence band and the Y2 point of the conduction band, and a direct band gap at $\Gamma$, where the energy of the direct gap is 0.4 eV larger than that of the indirect gap, as also found in experiments. The differences in the values of the experimental and theoretical energies (4.15 and 4.50 eV in experiments and 2.68 and 3.25 eV in calculations) are related to the well-known fact that PBE calculations tend to underestimate the band-gap energy.[51]

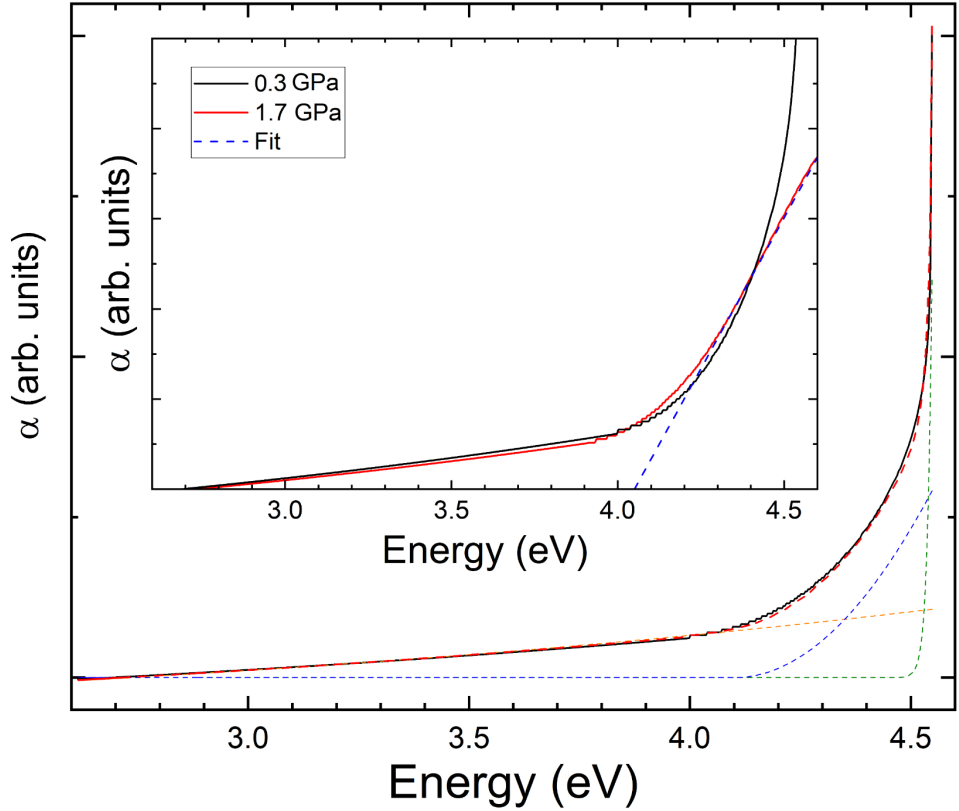

**Figure 9:** Absorbance ($\alpha$) of $K_2Zn(IO_3)_4 \cdot 2H_2O$ at 0.3 GPa (black line). The orange, blue, and green dashed lines are the contributions of the Urbach tail and indirect and direct gaps, respectively. The red dashed line is the fit including the three contributions. The inset shows the absorbance at 0.3 (black) and 1.7 GPa (red). The shape of the spectrum changes considerably between these pressures due to changes in the band structure as described in the text. The blue line is the fit used to determine the band-gap energy at 1.7 GPa.

Under compression, we noticed that beyond 0.70(5) GPa there was a change in the absorption spectrum wherein the high-energy component related to a direct band gap completely disappeared. This can be seen in the inset of Figure 9 where we compare the absorbances at 0.30(5) GPa and 1.70(5) GPa. The observed phenomenon is a consequence of changes in the topology of the band structure, as shown in Figure 10. Figure 10(b) shows the band structure at 1.6 GPa. It can be seen that because of the changes in the crystal structure, the band gap is reduced, and the top of the valence band moves from the $\Gamma$ point of the Brillouin zone to a point in between the $\Gamma$ and Y2 points. As a result, both the fundamental and the second band gaps are both indirect at 1.6 GPa. Such a change in the valence band causes, at 1.70(5) GPa, the change observed in the absorbance, and particularly the disappearance of the strong direct-gap absorption that was present at 0.30(5) GPa. At higher pressures, the band gap is further decreased, and the top of the valence band moves further towards the Y2 point of the Brillouin zone. This can be seen in Fig. 10(c) where we represent the band structure at 20 GPa.

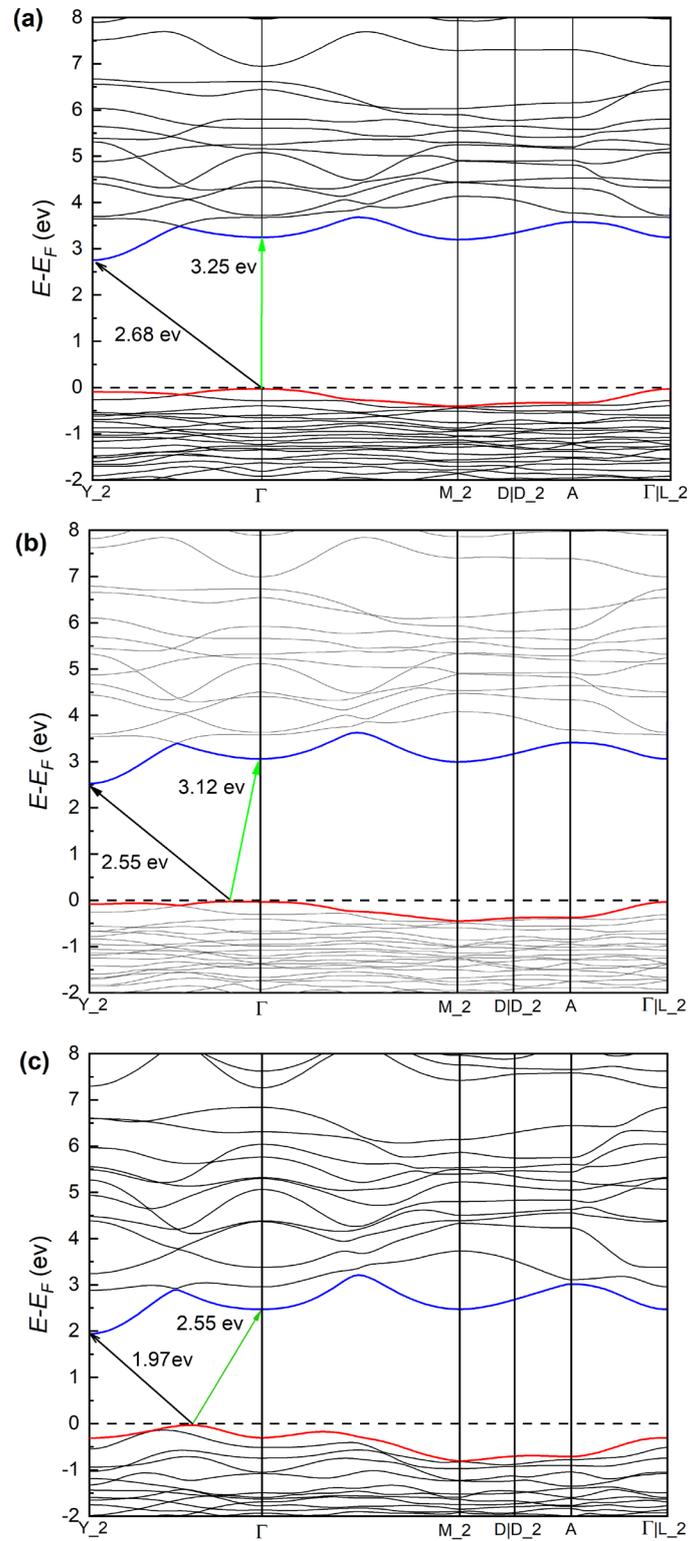

**Figure 10:** Calculated band structure of $K_2Zn(IO_3)_4 \cdot 2H_2O$ at 0 GPa (a), 1.6 GPa (b), and 20 GPa (c). The fundamental and second band gaps are identified by black and green arrows, respectively.

In Figure 11 we present results obtained from absorption experiments up to 20.10(5) GPa. In the experiments the absorbance of the sample redshifts, as found in the calculations. Additionally, the shape of the absorbance does not change supporting the hypothesis that at all pressures beyond 0.70(5) GPa the absorption is caused only by an indirect band gap and the low-energy Urbach absorption. From the experiments we obtained the pressure dependence of the band-gap energy. The results are shown in Figure 12 where they are compared with calculations. The results from density-functional theory have been offset 1.6 eV to facilitate the comparison of pressure dependences. Both methods give a similar pressure dependence up to 15 GPa. The band gap closes with a pressure coefficient of $dE_g/dP$ = -20(4) meV/GPa. A similar redshift of the band gap has been reported for $Mg(IO_3)_2$,[9] $Sr(IO_3)_2HIO_3$,[4] and $Na_3Bi(IO_3)_6$.[12] The behavior exhibited by the band gap of all these compounds results from the increase in length in the nearest neighbor iodine-oxygen bonds induced by compression.[52]

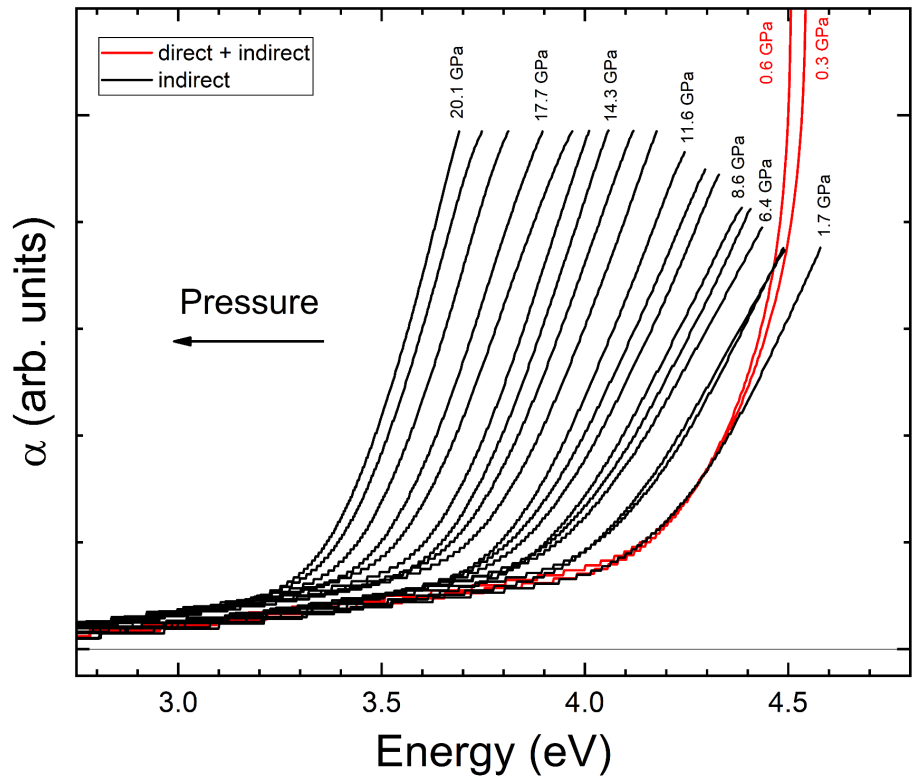

**Figure 11:** Absorbance ($\alpha$) of $K_2Zn(IO_3)_4 \cdot 2H_2O$ at different pressures. The spectra shown in red are from pressures where contributions from both the direct and indirect band gaps are present, while the spectra shown in black are from pressures where only the indirect band gap contributes.

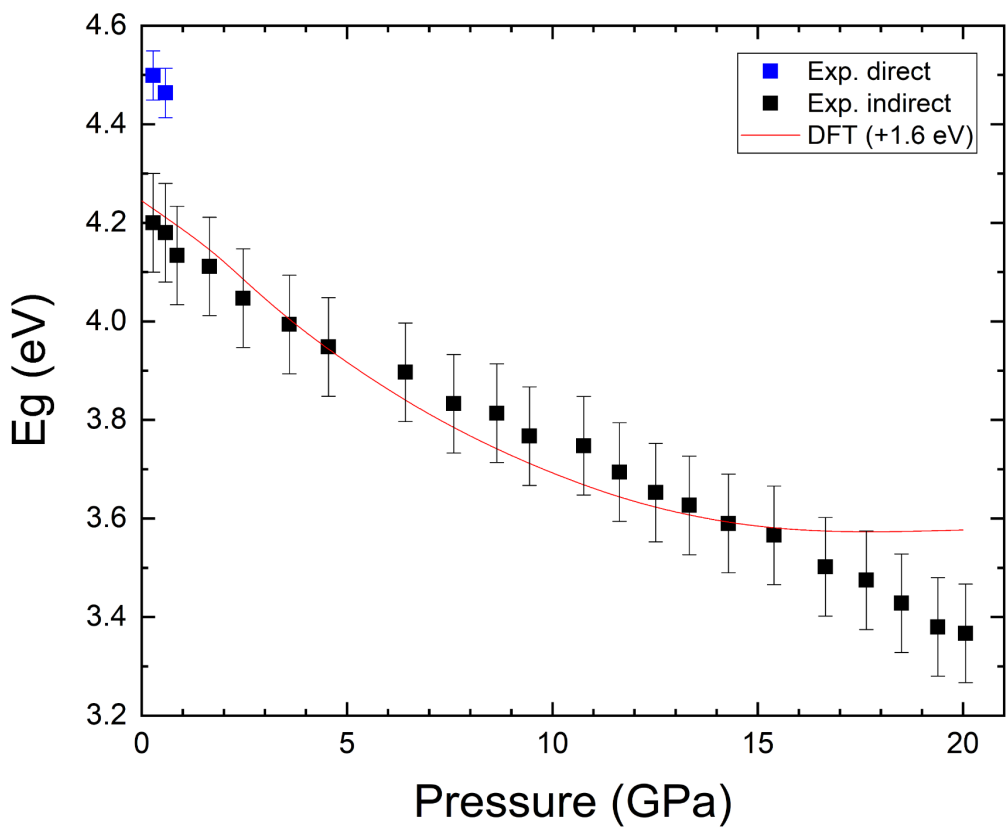

**Figure 12:** Pressure dependence of the band-gap energy ($E_g$) in $K_2Zn(IO_3)_4 \cdot 2H_2O$ as determined in experiment (symbols) and DFT calculations (line).

The behavior of the band-gap energy under compression can be understood using the electronic density of states (DOS) which is plotted in Figure 13. This figure shows that the valence band maximum of $K_2Zn(IO_3)_4 \cdot 2H_2O$ is primarily influenced by the 2p orbitals of O atoms, while the conduction band minimum is formed by contributions from both 2p and 5p orbitals of the O and I atoms, respectively. Thus, the band-gap energy is sensitive to iodine and oxygen interactions. It has been shown that the bandgap energy, $E_g$, in metal iodates can be correlated to the average I–O distance;[52] *i.e.* the larger the I–O distance, the smaller the bandgap energy, $E_g$. This is because the enlargement of I–O distances leads to a reduction in the hybridization of the 2p orbitals of O and the 5p orbitals of I, consequently diminishing the energy disparity between the bonding and antibonding states. As a result, the band-gap energy is expected to decrease. Our results are consistent with this interpretation. From 0 to 20 GPa the average I–O distance of the covalent bond within the $IO_3$ pyramid increases from 1.81(1) Å to 1.84(1) Å and the band gap decreases from 4.2(1) to 3.4(1) eV.

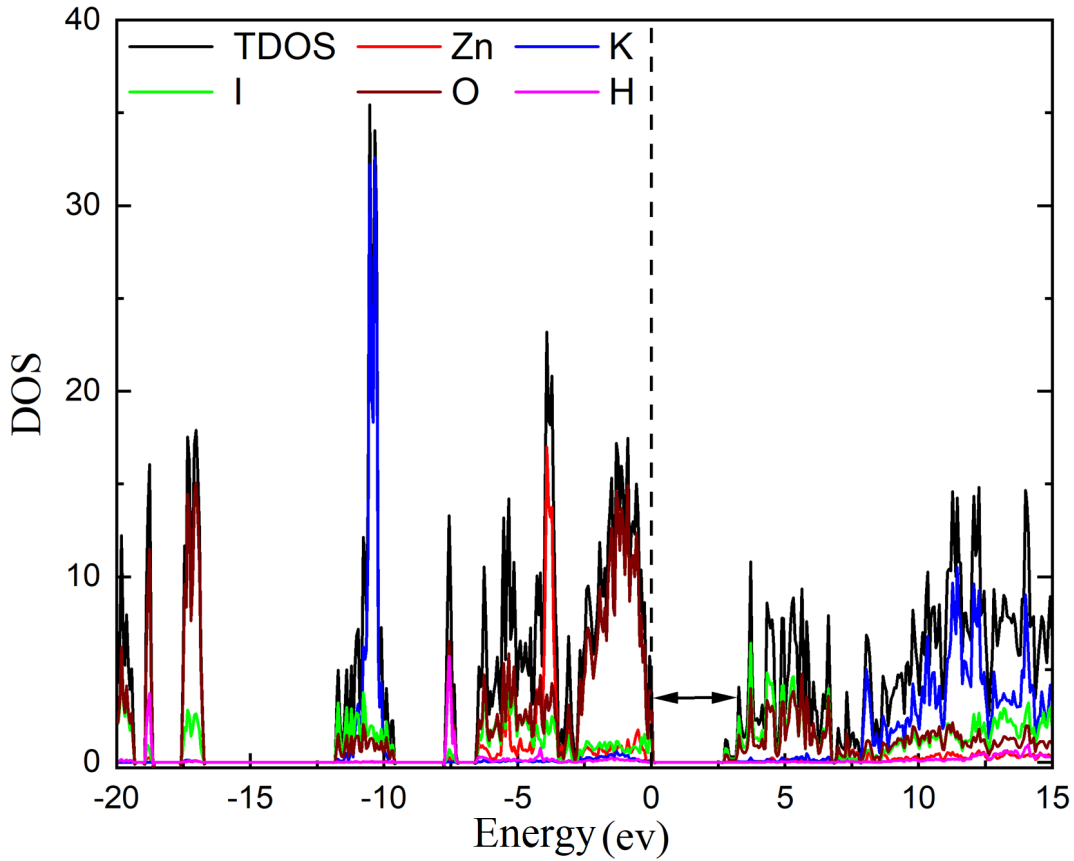

**Figure 13:** Calculated total and partial electronic density of states at 0 GPa.

## IV.  Conclusions

In this study, we present findings from synchrotron-based powder X-ray diffraction and optical absorption measurements conducted up to 20 GPa on $K_2Zn(IO_3)_4·2H_2O$. These experimental results have been integrated with density-functional theory calculations and a computational analysis of the electron-density topology. The compressibility of $K_2Zn(IO_3)_4·2H_2O$ has been examined, revealing that this material exhibits significant compressibility, notably as one of the most compressible iodates known. A key outcome of the research is the identification of pressure-induced configurational alterations that result in iodine hypercoordination and the formation of two-dimensional infinite iodate layers. We provide evidence indicating that pressure induces internal modifications in the crystal structure, promoting the transformation of covalent I–O bonds and halogen I···O interactions into O–I–O electron-deficient multicenter bonds. This chemical transformation influences the crystal structure and the electronic band structure. Notably, an intriguing phenomenon is the anomalous expansion of certain covalent I–O bonds involved in the formation of the pressure-induced electron-deficient multicenter bonds. Summing up, this study provides yet another example of a material which exhibits pressure-induced multicenter bonds. The expansion of the covalent I–O bonds leads to a large reduction in the band-gap energy, decreasing from 4.2(1) eV at ambient pressure to 3.4(1) eV at 20 GPa.

**Data Availability**

The data that support the findings of this study are available from the corresponding author upon reasonable request.

**Supporting Information**

Unit-cell parameters at different pressures and cif files with complete structural information at five selected pressures.

**AUTHOR INFORMATION**

**Corresponding Author**

**Daniel Errandonea** – Departamento de Física Aplicada-ICMUV-MALTA Consolider Team, Universitat de Valencia, 46100 Valencia, Spain; https://orcid.org/0000-0003-0189-4221; Email: daniel.errandonea@uv.es

Authors

**Robin Turnbull** – Departamento de Física Aplicada-ICMUV-MALTA Consolider Team, Universitat de Valencia, 46100 Valencia, Spain; https://orcid.org/0000-0001-7912-0248; Email: robin.turnbull@uv.es

**Hussien Helmy Hassan Osman** – Departamento de Física Aplicada-ICMUV-MALTA Consolider Team, Universitat de Valencia, 46100 Valencia, Spain and Instituto de Diseño para la Fabricación y Producción Automatizada, MALTA Consolider Team, Universitat Politècnica de València, 46022 València, Spain; https://orcid.org/0000-0003-1411-8299; Email: hussien.helmy@uv.es

**Zoulikha Hebboul** - Laboratoire Physico-Chimie des Matériaux, Université Amar Telidji de Laghouat, BP 37G, Route de Ghardaia, Laghouat 03000, Algeria; https://orcid.org/0000-0002-4443-8632; Email: z.hebboul@lagh-univ.dz

**Pablo Botella** - Departamento de Física Aplicada-ICMUV, MALTA Consolider Team, Universitat de Valencia, Valencia 46100, Spain; https://orcid.org/0000-0001-6930-8415; Email: pablo.botella-vives@uv.es

**Neha Bura** - Departamento de Física Aplicada-ICMUV, MALTA Consolider Team, Universitat de Valencia, Valencia 46100, Spain; Email: neha.bura@uv.es

**Peijie Zhang** - Departamento de Física Aplicada-ICMUV, MALTA Consolider Team, Universitat de Valencia, Valencia 46100, Spain; https://orcid.org/0000-0001-6355-5482; Email: Peijie.Zhang@uv.es

**Jose Luis Rodrigo-Ramon** - Departamento de Física Aplicada-ICMUV, MALTA Consolider Team, Universitat de Valencia, Valencia 46100, Spain; https://orcid.org/0009-0004-0485-238X; Email: Jose.L.Rodrigo@uv.es

**Josu Sanchez-Martin** - Departamento de Física Aplicada-ICMUV, MALTA Consolider Team, Universitat de Valencia, Valencia 46100, Spain; https://orcid.org/0000-0003-0241-0217; Email: Josu.Sanchez@uv.es

**Catalin Popescu** - CELLS-ALBA Synchrotron Light Facility, Cerdanyola 08290, Barcelona, Spain; https://orcid.org/0000-0001-6613-4739; Email: cpopescu@cells.es

**Francisco J. Manjón** - Instituto de Diseño para la Fabricación y Producción Automatizada, MALTA Consolider Team, Universitat Politècnica de València, Camí de Vera s/n, 46022 València, Spain; https://orcid.org/0000-0002-3926-1705; Email: fjmanjon@fis.upv.es

**Author Contributions**

D.E.: Conceptualization, formal analysis, writing–original draft, writing–review and editing. R.T.: Data acquisition, writing–review and editing. H.H.H.O, P.B., N.B., P.Z., J.L.R.R., J.S.M., C.P: methodology and writing–review and editing. Z.H.: Sample preparation and writing–review and editing. F.J.M: Conceptualization, formal analysis, writing–original draft, writing–review and editing. All authors have given approval to the final version of the manuscript.

**Notes**

The authors declare no competing financial interest.

**ACKNOWLEDGMENTS**

This work was financially supported by the Spanish Research Agency (AEI) and Spanish Ministry of Science, Innovation and Universities (MCIU) under grant PID2022-138076NB-C41/C42, and RED2022-134388-T, co-financed by EU FEDER funds (https://doi.org/10.13039/501100011033). This work was also supported by Generalitat Valenciana under Grant No. PROMETEO CIPROM/2021/075-GREENMAT, MFA/2022/007, and MFA/2022/025. This study forms part of the Advanced Materials program and is supported by MCIU with funding from the European Union Next Generation EU (PRTR-C17.I1) and by Generalitat Valenciana. D.E. and P.B. acknowledge funding from Generalitat Valenciana for the postdoctoral Fellowship No. CIAPOS/2023/406. R.T. acknowledges funding from Generalitat Valenciana for the postdoctoral Fellowship No. CIAPOS/2021/20 and Spanish Ministerio de Ciencia e Innovación (MICINN) and Agencia Estatal de Investigación (MCIN/AEI/ 10.13039/501100011033) under grant PID2021-125518NB-I00. J.S.-M. thanks the Spanish MCIU for the PRE2020-092198 fellowship. C. P. acknowledges the financial support from the Spanish MCIU through grant PID2021-125927NB-C21. The authors thank ALBA for providing beamtime under experiment no. 2023087668. Z.H. was supported by the Algerian Ministry of Higher Education and Scientific Research and is member of project "Projects de Recherche Formation-Universitaire" PRFU—PROJECT under contract B00L01UN030120220002.

For Table of Contents Only

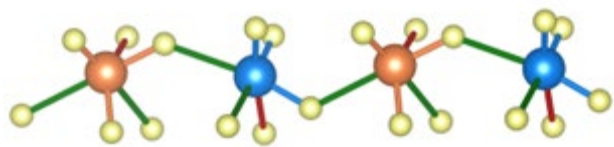